\documentclass[manuscript]{emulateapj}
\usepackage[usenames,dvips]{color}
\usepackage{graphicx,natbib}
\usepackage{multirow}
\usepackage{amssymb}
\usepackage{ulem}
\usepackage{epsfig}
\usepackage{amsmath}

\newcommand{\psr}{PSR~J1119$-$6127}

\shorttitle{HE emission from PSR J1119 after the 2016 outburst}
\shortauthors{Lin et al.}

\begin{document}

\title{Investigation of the High-Energy Emission from the Magnetar-like Pulsar PSR J1119$-$6127 after the 2016 Outburst 
     }

\author{Lupin Chun-Che Lin${}^1$,
         Hui-Hui Wang${}^2$,         
         Kwan-Lok Li${}^3$, 
         Jumpei Takata${}^2$,
         Chin-Ping Hu${}^4$,
         C.-Y. Ng${}^4$, 
         C. Y. Hui${}^5$, 
         A. K. H. Kong${}^6$,
         Pak-Hin T. Tam${}^7$ and 
         Paul K. H. Yeung${}^8$
         }
 \affil{${}^1$
        Department of Physics, UNIST, 
        Ulsan 44919, Korea;
        lupin@unist.ac.kr}
\affil{${}^2$
       School of Physics, 
       Huazhong University of Science and Technology, 
       Wuhan 430074, China.}
 \affil{${}^3$
       Department of Physics and Astronomy, 
       Michigan State University, East Lansing, 
       MI 48824, USA}
 \affil{${}^4$
       Department of Physics, 
       The University of Hong Kong, Pokfulam Road, 
       Hong Kong} 
 \affil{${}^5$
       Department of Astronomy and Space Science, 
       Chungnam National University, 
       Daejeon 305-764, Korea}  
\affil{${}^6$
	    Institute of Astronomy, 
	    National Tsing Hua University, Hsinchu, 
	    Taiwan} 
\affil{${}^7$      
		School of Physics and Astronomy, 
		Sun Yat-sen University, Zhuhai 519082, China}     
\affil{${}^8$      
       Institute of Experimental Physics, Department of Physics, 
       University of Hamburg, 
       Luruper Chaussee 149, D-22761 Hamburg, Germany}

\begin{abstract}
\psr\ is a radio pulsar that behaved magnetar like bursts, and we performed a comprehensive investigation of this pulsar using the archival high-energy observations obtained after its outburst in 2016 July. 
After the 2016 outburst, specific regions on the neutron star surface were heated up to $> 0.3$ and $> 1 $~keV from $\sim$0.2~keV. 
A hard non-thermal spectral component with a photon index $<$ 0.5 related to the magnetospheric emission can be resolved from the {\it NuSTAR} spectra above 10 keV.
We find that the thermal emitting regions did not cool down and gradually shrank by about 20--35\% four months after the outburst.
Hard X-ray pulsations were detected with {\it NuSTAR} immediately after the outburst, at 5$\sigma$ confidence level and with a background-subtracted pulsed fraction of $40\pm10$\%. 
However, the signal became undetectable after a few days.
Using {\it Fermi} data, we found that the gamma-ray emission in 0.5--300 GeV was suppressed along with the disappearance of the radio pulsations. 
This is likely caused by a reconfiguration of the magnetic field.
We also discovered that the timing noise evolved dramatically, and the spin-down rate significantly increased after the 2016 glitch.
We proposed that post-outburst temporal and spectral behaviors from radio to gamma-ray bands were caused by changes of the magnetosphere structure, pair plasma injection and the shrinking emission sites on the neutron star.

\end{abstract}


\section{Introduction}
\label{sec:intro}
Magnetars are a special class of pulsars that have relatively long spin periods and extremely strong surface magnetic field of $10^{14}$--$10^{15}$\,G \citep{Kaspi2017} although some of them can have relatively weak surface fields (e.g., $\sim 6\times 10^{12}$~G for SGR 0418+5729; \citealt{Rea2013}). 
The soft X-ray emission of magnetars is commonly believed to be powered by the decay of the magnetic field \citep{TD96}. 
One of the most intriguing features for magnetars is bursts in hard X-rays/soft $\gamma$-rays \citep{WT2006,Ng2011,PR2012} when the local magnetic stress becomes too strong to be balanced by the restoring force of the crust or the reconnection of the magnetic field leads to the change of the magnetosphere structure.  
Some rotation-powered pulsars (RPPs) with a high surface magnetic field strength ($\gtrsim 4\times 10^{13}$~G; hereafter, high-$B$ pulsars) show similar outburst activities and are now considered as the transitional objects between canonical RPPs and magnetars \citep{KP2005,NKl2011,Hu2017}.
High-$B$ pulsars and magnetars also share similar features in their outburst activities, glitch sizes and the inferred level of timing noise. 
Studies of high-$B$ pulsars is essential to unify the connection between magnetars and isolated neutron stars (NS)/radio pulsars \citep{PG2007,PP2011,Vigano2013}.  
Most high-$B$ pulsars were detected in the radio band (e.g., PSRs J1718$-$3718, J1814$-$1733, J1814$-$1733: \citealt{Hobbs2004,Camilo2000,McLaughLin2003}) and a few of them are radio-quiet (e.g., PSR J1846$-$0258: \citealt{AKLM2008,Gavriil2008}).
Similar to some transient magnetars (e.g., 1E 1547.0$-$5408; \citealt{Burgay2009}), intermittent and variable radio pulses were detected from some high-$B$ pulsars \citep{Thompson2008}.
Among this class of objects, \psr\ is a particularly interesting case because it has a confirmed rotational period of $\sim$408~ms detected in the radio \citep{Camilo2000}, X-ray \citep{GKGP2005} and $\gamma$-ray \citep{Parent2011} bands, which permits a rich, multiwavelength investigation. 

\psr\ is associated with the supernova remnant (SNR) G292.2$-$0.5 \citep{Pivovaroff2001,Ng2012}, and its pulsar wind nebula (PWN) was resolved by {\it Chandra} \citep{GS2003}. 
Based on the measurements through H{\sc i} absorption, the distance of the pulsar is inferred to be 8.4~kpc \citep{CMC2004}.   
According to the detected spin down rate $\dot{P}=4.0\times 10^{-12}$ \citep{Camilo2000}, the inferred surface magnetic field is $4.1\times 10^{13}$~G and the spin-down luminosity is $\dot{E}=2.3\times 10^{36}$~erg s$^{-1}$. 
The characteristic age was estimated as 1600\,yr with the breaking index $2.684\pm0.002$ determined from radio timing data over 12 years \citep{WJE2011}.
\psr\ experienced three glitches in 1999, 2004 and 2007 \citep{Camilo2000,WJE2011,Anton2015}, but no outburst event was detected before.
Recoveries were observed following all these glitches \citep{WJE2011}, similar to that of many pulsars and magnetars \citep{DKG2008}.

The Gamma-Ray Burst Monitor onbroad the {\it Fermi Gamma-ray Space Telescope} was triggered by a magnetar-like burst of \psr\ on 2016 July 27 \citep{Younes2016}, and {\it Neil Gehrels Swift Observatory} (hereafter {\it Swift}) Burst Alert Telescope also reported a similar burst after a few hours \citep{Kennea2016}. 
The known X-ray pulsations were subsequently confirmed by the X-Ray Telescope (XRT) onboard {\it Swift} \citep{AVE2016}, and several investigations were performed to study its timing and spectral behavior in the post-outburst stage \citep{ATSK2016,Majid2016}. 
Based on the detailed examination with S/N-based search and Bayesian blocks algorithm, 13 short X-ray bursts were identified between 2016 July 26 and 28 \citep{Gogus2016}. 
The pulsar showed spin-up glitch of $\Delta \nu=1.40(2)\times 10^{-5}$~Hz and  $\Delta \dot{\nu}=-1.9(5)\times 10^{-12}$~Hz s$^{-1}$ soon after the onset of 2016 outburst (\citealt{Archibald2016}; hereafter, after the 2016 outburst).
The initial follow-up radio observations did not detect any pulsations, but the signal re-emerged over a large frequency band with different pulse profiles in 2016 mid-Aug \citep{Majid2017}.
Such a transient radio emission feature could be associated with the glitching activity as seen from magnetars \citep{Camilo2006}.
In addition, there was a significant difference in the radio pulse profile shape before and after the outburst, from single peaked to a complex profile with multiple components at 1.4~GHz \citep{Archibald2017}.
All the observed features, including the outburst behavior and the unusual change of the radio profiles, are similar to those of transient magnetars.

The X-ray spectrum of \psr\ consists of thermal and non-thermal components in the quiescent state,  \citep{GKGP2005,SK2008,Ng2012}. 
The thermal emission can be attributed to a hotspot with $kT \sim 0.21$\,keV or NS atmospheric emission with $kT=0.06$--0.17\,keV, and the non-thermal component can be described by a simple power-law (PL) with a photon index ($\Gamma$) of 1.6--2.1.
The $\gamma$-ray spectrum can be characterized with a PL with an exponential cutoff (PLEC), similar to other RPPs \citep{Parent2011}.  
After the 2016 outburst, the X-ray spectrum was dominated by a thermal component, and the PL component became harder with a photon index of $1.2\pm 0.2$, extending to the hard X-ray band \citep{Archibald2016}.
The 0.5--10~keV luminosity is $3.5 \times 10^{35}$~erg~s$^{-1}$, corresponding to $\sim$15\% of the pulsar spin-down power and is comparable to the pre-outburst $\gamma$-ray ($>0.1$\,GeV) luminosity \citep{Parent2011}.  
At the end of 2016 Oct,  a single PL with a softer photon index of $2.0\pm0.2$ was sufficient to model the X-ray spectrum \citep{BSM2017}.

In this paper, we analyze the publicly available high-energy observations taken after the outburst.
The observations are described in Section~\ref{sec:observations}. 
We search for spin periodicities via data counting statistics at different epochs.
Then we trace the spin period evolution with phase-coherent analysis.  
We obtained the X- and $\gamma$-ray spectra before and after the outburst, and performed a quantitative comparison.
The results are presented in Section~\ref{sec:results}.
Based on our results, we present a coherent picture and discuss emission scenarios in Section~\ref{sec:discussion}.
A global magnetosphere reconfiguration could serve as the most possible origin.
Our contribution and two recent published articles including the long-term investigations of \psr\ in the radio \citep{Dai2018} and X-ray bands \citep{Archibald2018} can certainly provide a better understanding to the outburst/glitch event that occurred in this pulsar and further clarify its connection to transient radio RPPs and magnetars.

\section{High-energy Observations}
\label{sec:observations}
In order to characterize the high-energy emission of \psr, we analyzed the archival data obtained from {\it Swift}, {\it XMM-Newton}, Nuclear Spectroscopic Telescope Array ({\it NuSTAR}) and {\it Fermi} observatories. 
The pulsar position at RA=$11^{\rm h}19^{\rm m}14\fs{26}$, dec.=$-61^{\circ}27'49\farcs3$ (J2000), which was obtained from {\it Chandra} \citep{SK2008,BSM2017}, to perform the timing analysis of X-ray observations. 
All the photon arrival times were corrected to the barycentric dynamical time (TDB) with the JPL DE200 solar system ephemeris at the source position determined by {\it Chandra} \citep{SK2008} for all the datasets. 

\begin{table*}
\caption{{\small Spin period of \psr\ determined from the X-ray data after 2016 outburst.}}\label{spin}
\begin{tabular}{cccccclcc} 
\hline \hline {\footnotesize Start obs. date} & {\footnotesize Instruments} & {\footnotesize Duration} & {\footnotesize ObsID} & {\footnotesize Photons} & {\footnotesize Epoch zero$^a$} & {\footnotesize Spin frequency$^b$} & {\small $Z_1^2$/$H$} & {\footnotesize Chance}  \\
 &  & {\footnotesize (ks)} & & & {\footnotesize (MJD)} & {\footnotesize (Hz)} &  & {\footnotesize prob.} 
\\ \hline 
{\small 2016-07-28*} & {\footnotesize {\sl Swift}/XRT} & {\small 35.4} & {\footnotesize \dataset[http://www.swift.ac.uk/archive/browsedata.php?oid=00034632001\&source=obs]{00034632001}} & {\small 3846}  & {\small 57598.0}  & {\small 2.439842(1)}  & {\small 503/571}  & {\footnotesize $<10^{-11}$} 
\\ {\small 2016-07-28*} & {\footnotesize {\sl NuSTAR}/FPM} & {\small 82.9} & {\footnotesize \dataset[https://heasarc.gsfc.nasa.gov/FTP/nustar/data/obs/01/8/80102048002]{80102048002}} & {\small 32265}  & {\small 57598.5}  & {\small 2.4398409(2)}  & {\small 5820/6980}  & {\footnotesize $<10^{-11}$} 
\\ {\small 2016-07-31*} & {\footnotesize {\sl Swift}/XRT} & {\small 17.9} & {\footnotesize \dataset[http://www.swift.ac.uk/archive/browsedata.php?oid=00034632002\&source=obs]{00034632002}} & {\small 424}  & {\small 57600.2}  & {\small 2.43983(2)}  & {\small 28.6/47.6}  & {\footnotesize $5.4\times 10^{-7}$} 
\\ {\small 2016-08-05} & {\footnotesize {\sl NuSTAR}/FPM} & {\small 127.0} & {\footnotesize \dataset[https://heasarc.gsfc.nasa.gov/FTP/nustar/data/obs/01/8/80102048004]{80102048004}} & {\small 40275}  & {\small 57606.0}  & {\small 2.4398219(1)}  & {\small 7670/9310}  & {\footnotesize $<10^{-11}$} 
\\ {\small 2016-08-06} & {\footnotesize {\sl XMM}/pn} & {\small 20.1} & {\footnotesize \dataset[http://nxsa.esac.esa.int/nxsa-web/\#obsid=0741732601]{0741732601}} & {\small 25530}  & {\small 57606.6}  & {\small 2.4398214(8)}  & {\small 7070/7990}  & {\footnotesize $<10^{-11}$} 
\\ {\small 2016-08-09} & {\footnotesize {\sl Swift}/XRT} & {\small 57.6} & {\footnotesize \dataset[http://www.swift.ac.uk/archive/browsedata.php?oid=00034632007\&source=obs]{00034632007}} & {\small 1376}  & {\small 57609.5}  & {\small 2.439814(2)}  & {\small 151/177}  & {\footnotesize $<10^{-11}$} 
\\ {\small 2016-08-10} & {\footnotesize {\sl Swift}/XRT} & {\small 5.9} & {\footnotesize \dataset[http://www.swift.ac.uk/archive/browsedata.php?oid=00034632008\&source=obs]{00034632008}} & {\small 339}  & {\small 57610.25}  & {\small 2.43979(3)}  & {\small 64.5/79.9}  & {\footnotesize $<10^{-11}$} 
\\ {\small 2016-08-14} & {\footnotesize {\sl NuSTAR}/FPM} & {\small 170.8} & {\footnotesize \dataset[https://heasarc.gsfc.nasa.gov/FTP/nustar/data/obs/01/8/80102048006]{80102048006}} & {\small 32473}  & {\small 57615.0}  & {\small 2.4397973(1)}  & {\small 5340/6320}  & {\footnotesize $<10^{-11}$} 
\\ {\small 2016-08-15} & {\footnotesize {\sl XMM}/pn} & {\small 27.9} & {\footnotesize \dataset[http://nxsa.esac.esa.int/nxsa-web/\#obsid=0741732701]{0741732701}} & {\small 26060}  & {\small 57616.0}  & {\small 2.4397945(6)}  & {\small 7020/7850}  & {\footnotesize $<10^{-11}$}
\\ {\small 2016-08-26} & {\footnotesize {\sl Swift}/XRT} & {\small 70.5} & {\footnotesize \dataset[http://www.swift.ac.uk/archive/browsedata.php?oid=00034632010\&source=obs]{00034632010}} & {\small 627}  & {\small 57627.0}  & {\small 2.439746(3)}  & {\small 49.4/54.1}  & {\footnotesize $4.0\times 10^{-8}$}
\\ {\small 2016-08-30} & {\footnotesize {\sl XMM}/pn} & {\small 32.5} & {\footnotesize \dataset[http://nxsa.esac.esa.int/nxsa-web/\#obsid=0741732801]{0741732801}} & {\small 19466}  & {\small 57630.3}  & {\small 2.4397218(6)}  & {\small 4960/5540}  & {\footnotesize $<10^{-11}$}
\\ {\small 2016-08-30} & {\footnotesize {\sl Swift}/XRT} & {\small 17.9} & {\footnotesize \dataset[http://www.swift.ac.uk/archive/browsedata.php?oid=00034632011\&source=obs]{00034632011}} & {\small 542}  & {\small 57630.4}  & {\small 2.439720(8)}  & {\small 78.1/105}  & {\footnotesize $<10^{-11}$}
\\ {\small 2016-08-30} & {\footnotesize {\sl NuSTAR}/FPM} & {\small 166.5} & {\footnotesize \dataset[https://heasarc.gsfc.nasa.gov/FTP/nustar/data/obs/01/8/80102048008]{80102048008}} & {\small 20878}  & {\small 57631.0}  & {\small 2.4397154(2)}  & {\small 1810/1880}  & {\footnotesize $<10^{-11}$}
\\ {\small 2016-09-27} & {\footnotesize {\sl Swift}/XRT} & {\small 6.4} & {\footnotesize \dataset[http://www.swift.ac.uk/archive/browsedata.php?oid=00034632020\&source=obs]{00034632020}} & {\small 527}  & {\small 57658.05}  & {\small 2.43964(2)}  & {\small 72.7/99.9}  & {\footnotesize $<10^{-11}$}
\\ {\small 2016-12-12} & {\footnotesize {\sl NuSTAR}/FPM} & {\small 183.3} & {\footnotesize \dataset[https://heasarc.gsfc.nasa.gov/FTP/nustar/data/obs/01/8/80102048010]{80102048010}} & {\small 2118}  & {\small 57735.0}  & {\small 2.4391775(8)}  & {\small 175/203}  & {\footnotesize $<10^{-11}$}
\\ {\small 2016-12-13} & {\footnotesize {\sl XMM}/pn} & {\small 47.5} & {\footnotesize \dataset[http://nxsa.esac.esa.int/nxsa-web/\#obsid=0762032801]{0762032801}} & {\small 4053}  & {\small 57735.5}  & {\small 2.439177(1)}  & {\small 954/1020}  & {\footnotesize $<10^{-11}$}
\\
\hline 
\hline
\multicolumn{9}{l}{{\footnotesize $^{a}$ The time zero determined in the periodicity search.}} \\
\multicolumn{9}{l}{{\footnotesize $^{b}$ The uncertainties quoted in parentheses were assessed by Eq.~6a in \citet{Leahy87}.}}\\
\multicolumn{9}{l}{{\footnotesize $^{*}$ Data sets also used by \citet{Archibald2016} to generate the timing ephemeris.}
} 
\end{tabular}

\end{table*}


\subsection{{\sl Fermi}}
\label{ssec:Fermi}
We obtained the {\it Fermi} Large Area Telescope (LAT) Pass 8 (P8R2) data in the energy range of 0.5-300 GeV.
This range was chosen to minimize avoid the serious contamination by other sources in the source-crowded region \citep{Parent2011}. 
We considered the LAT data collected within 2013 November (MJD $\sim 56,600$)--2016 December (MJD $\sim 57,750$).
Photons in a circular region of interest within $20^{\circ}$ and $1^{\circ}$ from the position of 3FGL~J1119.1$-$6127, corresponding to \psr, determined in the LAT 4-year point source catalog \citep{3FGL} are taken into account for spectral and temporal analysis, respectively.
The data reduction was processed with {\it Fermi} Science Tools(v10r0p5).

We selected events in the class for the point source or Galactic diffuse analysis (i.e., event class 128) and considered photons collected in the front- and back-sections of the tracker (i.e., evttype = 3).
We used the instrument response function ``P8R2\_SOURCE\_V6'' throughout this study.
We also excluded events with zenith angles larger than $90^{\circ}$ to reduce contamination from Earth's albedo $\gamma$-rays, and the data quality is constrained by the good-time-interval of the spacecraft (i.e., DATA\_QUAL $>$ 0).
In the timing analysis, all the photon arrival times were corrected to TDB with the task of {\it gtbary} supported under {\it Fermi} Science tools. 

\subsection{{\sl Neil Gehrels Swift}}
\label{ssec:Swift}
We considered all the XRT observations operated under the windowed timing (WT) mode to perform the timing analysis, since only this mode provides a high enough time resolution of 1.76\,ms.
All the data were obtained from the HEASARC\footnote{https://heasarc.gsfc.nasa.gov/docs/archive.html} archive, and we obtained the light curves and spectra using the XRTGRBLC task in the HEASOFT package (v.6.22).
We extracted the source counts from a box region of $15''\times35''$, which is the default setting of the XRTGRBLC task. 
Only photon energies in the range of 0.3--10 keV with grades 0--2 were included for further analysis.
We corrected the photon arrival times to TDB using the HEASOFT task of {\it barycorr}.

\subsection{{\sl XMM-Newton}}
\label{ssec:XMM} 
After the outburst, {\it XMM-Newton} observed \psr\ on August 6, 15, 30 and December 13 of 2016, with exposure of $\sim$21.6\,ks, 29.5\,ks, 34\,ks and 49\,ks, respectively. 
All the EPIC cameras were operated in the small window mode with the thin filter. 
We considered  single- to quadruple-pixel events (PATTERN=0--12) for MOS and single and double events (PATTERN=0--4) for PN.
All the data reduction was performed using the latest XMM-Newton Science Analysis Software (XMMSAS version 16.1.0).
We also filtered out artifacts from the calibrated and concatenated dataset and put the most conservative events screening criteria as ``FLAG==0''. 
We removed all the photons collected in short bursts (e.g., \citealt{Archibald2017}) and time intervals contaminated by X-ray background flares in both timing and spectral analyses.
We extracted events in 0.15--12\,keV from a 20\arcsec\ radius region centered on the {\it Chandra} position. 
Note that this radius encircles $\sim76$\% of the energy.
We corrected the photon arrival times to TDB using the XMMSAS task {\it barycen} for the following timing analysis.
In the spectral analysis, we determined the background in the nearby source-free region and generated the response matrices and ancillary response files with the XMMSAS tasks {\it rmfgen} and {\it arfgen}. 

\subsection{{\sl NuSTAR}}
\label{ssec:NuSTAR}
Non-thermal X-rays related to \psr\ can have higher energy after bursts, so we also included the {\it NuSTAR} observations, which has an effective energy range in 3--79 keV \citep{Harr2013}. 
{\it NuSTAR} started to observe \psr\ following the main magnetar-like outburst on 2016 July 28.
Each data of {\it NuSTAR} was observed with the onboard focal plane modules A and B (FPMA/B).
Excluding two slew observations and some with exposure less than 1\,ks, we have five observations of 2016 July 28, Aug 5, 14, 30 and Dec 12, corresponding to exposure of $\sim$54.4\,ks, 87.2\,ks, 95.4\,ks, 92.1\,ks and 94.3\,ks, respectively.
We carried out detailed temporal and spectral analysis on the later four datasets, which were jointly observed with {\it XMM-Newton}. 
We only performed the temporal analysis on the first dataset (ObsID 80102048002) since the joint spectral fit with the {\it Swift} observation has previously been reported \citep{Archibald2016}.

We also used the HEASOFT package (v.6.22) together with NuSTARDAS v1.8.0 and the updated {\it NuSTAR} calibration database (CALDB version 20171204) to for the data analysis.
We adopted a source region of $60^{\prime\prime}$ radius (and $30^{\prime\prime}$ for the last observation, during which the source was faint) and a source-free background region with a radius of $90^{\prime\prime}$. 
We extracted events with the pulse-invariant channel of 35--1909, which corresponds to the default energy range of 3--79 keV and generated the response matrices using the {\it nuproducts} tasks.
Similar to the {\it Swift} data, we also used the {\it barycorr} task to perform the barycentric time correction. 

\begin{figure}[t]
  \includegraphics[angle=0,scale=0.36]{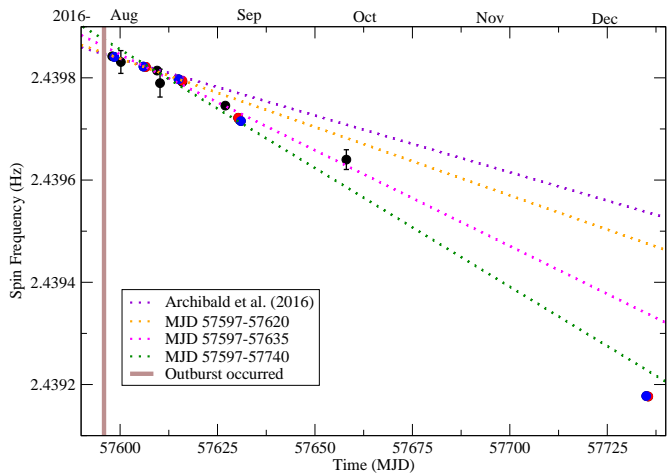} 
\caption{\footnotesize{
Evolution of the frequency of \psr determined from the archival X-ray data after the 2016 magnetar-like outburst.
The black, red and blue data points represent the best-fit frequency obtained from {\it Swift}/XRT, {\it XMM}/PN and {\it NuSTAR}/FPM, respectively. 
The brown solid line labels the onset of 2016 outburst.
The dotted lines denote the linear regression over different time intervals.
The magenta dotted lines describe the post-outburst ephemeris reported by \citet{Archibald2016} with spin frequency ($\nu$) = 2.43983734(8)~s$^{-1}$ and the spin-down rate ($\dot{\nu}$) = $-2.57(5)\times 10^{-11}$~s$^{-2}$ at the epoch of MJD 57,600.
At the same epoch zero, orange, pink and green lines correspond to $\dot{\nu}$ of $-3.10(3)\times 10^{-11}$, $-4.3(3)\times 10^{-11}$ and $-5.4(2)\times 10^{-11}$~s$^{-2}$, respectively.
}
\label{fig:evo_spin}
}
 \end{figure}
\begin{figure}[bp]
\centering
\hspace*{\fill}{\includegraphics[scale=0.355]{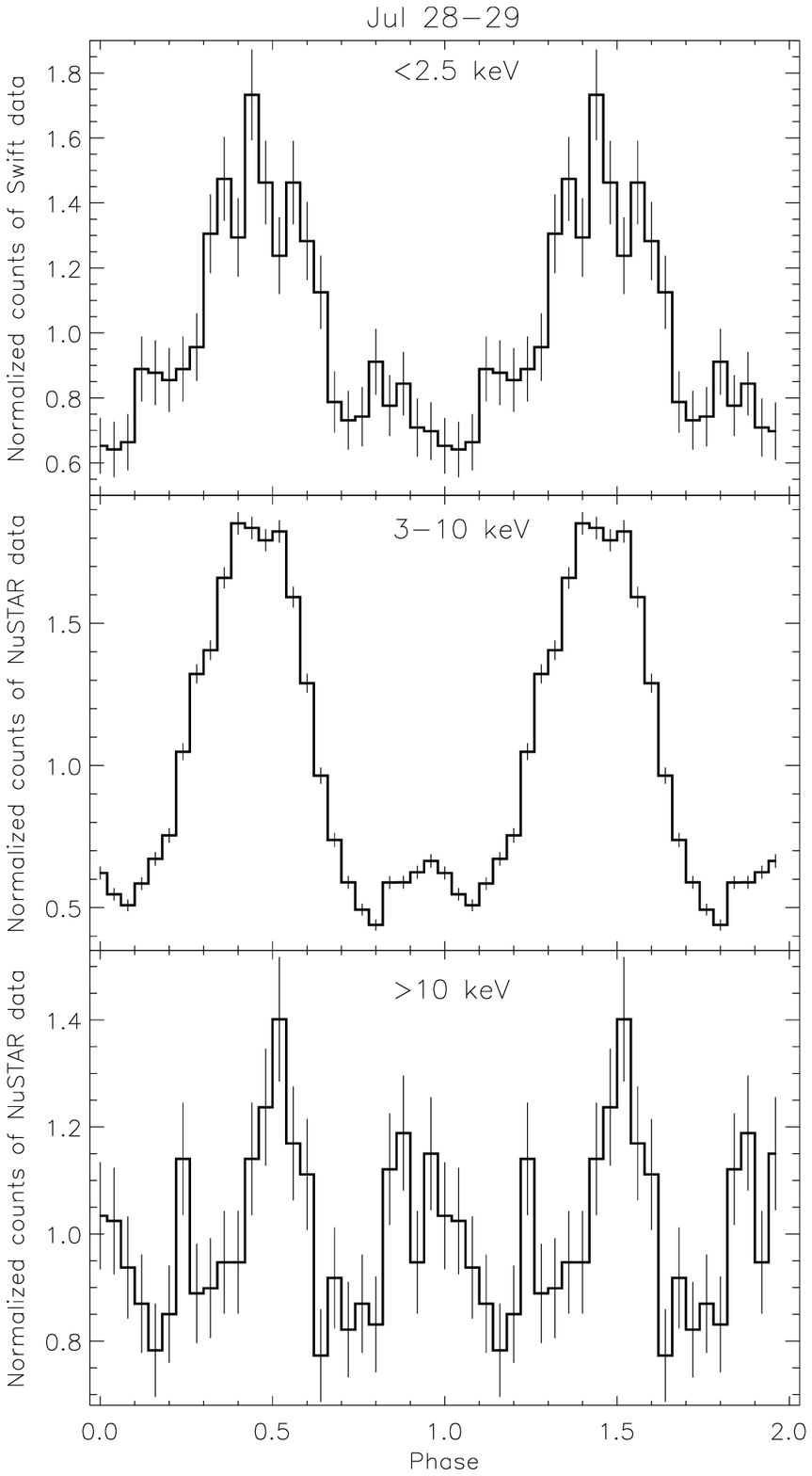}}
\hspace*{\fill}{\includegraphics[scale=0.355]{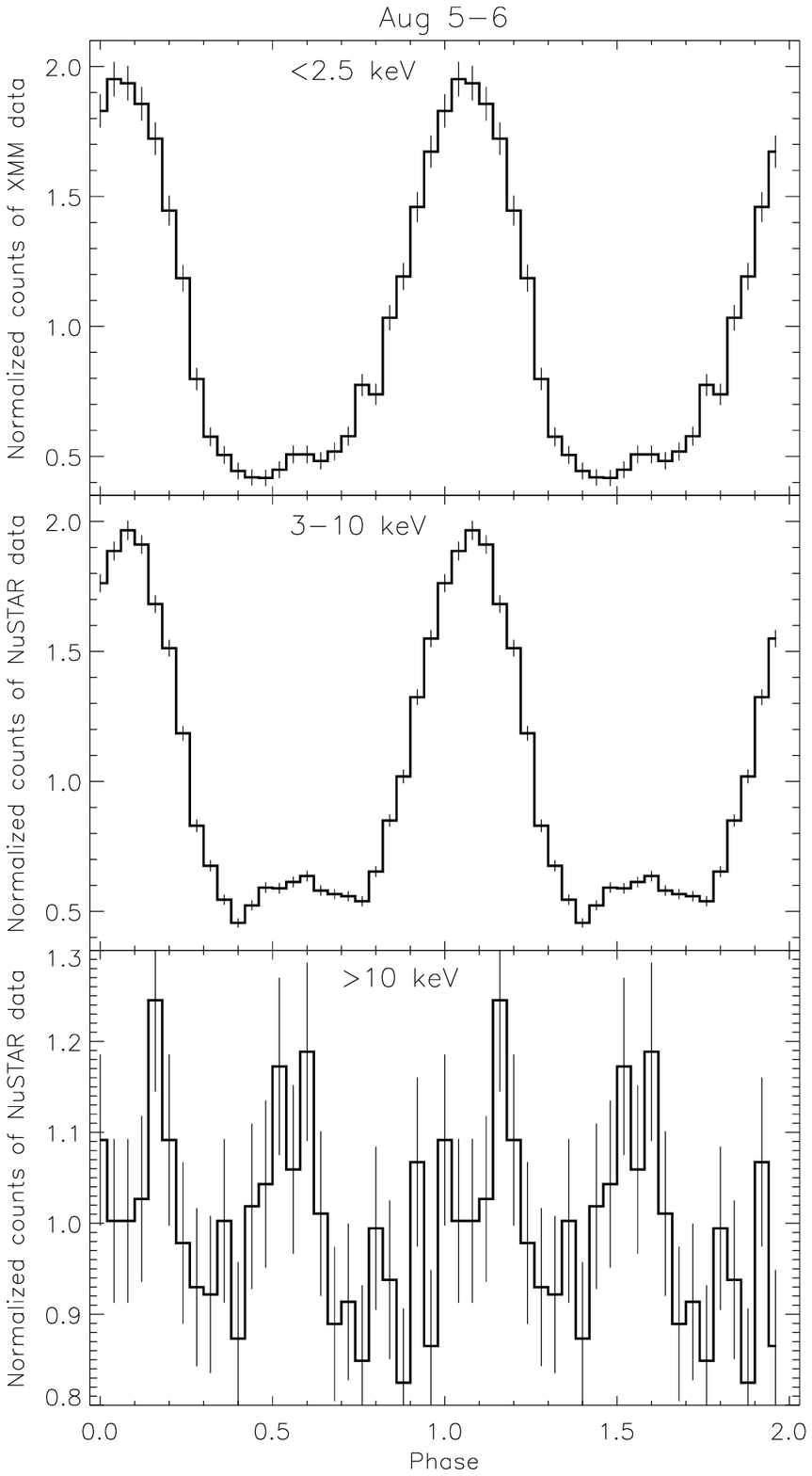}}
\hspace*{\fill}
\caption{\footnotesize{Post-outburst X-ray pulse profiles of \psr\ taken on July 28--29 (left panel) and Aug 5--6 (right panel). The soft X-ray band profiles are obtained from {\it Swift}/XRT and {\it XMM}/PN, respectively, and the medium and hard X-ray profiles are from {\it NuSTAR}/FPM. Two cycles of each profile are shown for clarity.}}
\label{fig:ERFLC}
\end{figure}
 
\section{Results}
\label{sec:results}
We analyzed the X-ray and $\gamma$-ray data of \psr mentioned in section~\ref{sec:observations} and inspected the spin periodicity and spectral behavior of our target at different epochs.
In the following subsections, we will present the detailed procedures in our analyses and the complete results.

\subsection{{\rm X-Ray Timing Analysis}}
\label{ssec:Timing}
We only utilized the data obtained with the PN camera for {\it XMM-Newton} observations to perform timing analysis because of its sufficient temporal resolution (5.7 ms).
We extrapolated the pre-outburst timing solution to search for the periodic signals among all the X-ray data.
We performed 100 trials with a step size determined by the Fourier width (i.e., 1/time span of the data).
Only spin frequencies with a chance probability of the signal less than $5.7\times 10^{-7}$ (i.e., 5$\sigma$ confidence level) yielded from $H$-statistics \citep{JB2010} are presented in Table~\ref{spin}.
Because the pulse profiles are mostly single peaked, all the results also correspond to a significant Rayleigh power \citep{Mardia72,Gibson82}, except for Obs ID. 00034632002 investigated by {\it Swift}.  
The evolution of the spin frequency, detected in X-ray data, is shown in Fig.~\ref{fig:evo_spin}, and the spin-down rate drastically increased after the end of 2016 July.

\begin{figure}
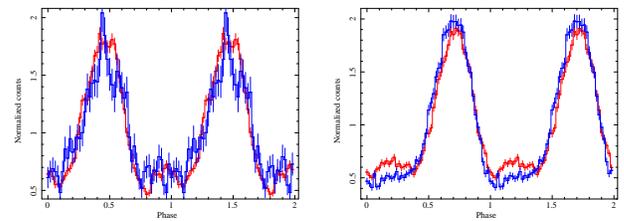

\centering
\hspace*{\fill}{\includegraphics[angle=-90,scale=0.16]{cmp1.eps}}
\hspace*{\fill}{\includegraphics[angle=-90,scale=0.16]{cmp2.eps}}
\hspace*{\fill}
\caption{\footnotesize{Comparison of the pulse profiles obtained from the different instruments. The left panel shows {\it Swift}/XRT (blue line) vs. {\it NuSTAR}/FPM (red line), and the right panel shows {\it XMM}/PN (blue line) vs. {\it NuSTAR}/FPM (red line). Two cycles of each profile are shown for clarity.}}
\label{fig:cmp_FLC}
\end{figure}
\begin{figure}[bp]
  \includegraphics[angle=0,scale=0.36]{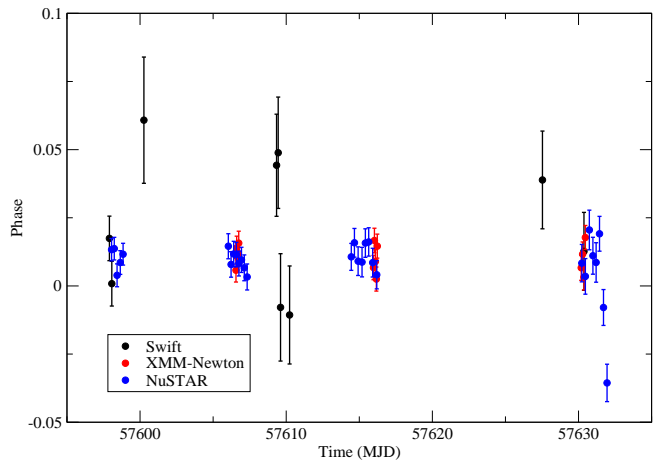} 
\caption{\footnotesize{Residuals for the post-outburst timing solution of MJD 57,598--57,632.  
}
\label{fig:residual}
}
 \end{figure}

Fig.~\ref{fig:ERFLC} shows the folded light curves in the different energy bands.
In the soft X-ray (0.15 or 0.3$-2.5$~keV) and medium (3--10~keV) bands, there is no significant structural change with aligned pulsed peaks in the pulse profile, consistent with that reported by \citet{Archibald2016}.
In the hard X-ray band (10--79\,keV), we found a double-peaked pulse profile on July 28--29 with a false alert probability of $7.7\times 10^{-7}$ through $H$-statistics.
Nevertheless, the hard X-ray pulsations became much weaker after Aug and were only detected at less than 3$\sigma$ level.  

We traced the Times-of-Arrivals (TOAs; \citealt{Ray2011}) inferred from the positive detections mentioned in Table~\ref{spin}, using a maximum likelihood method \citep{Livingstone2009}, to account for the influence of timing noise.
Because the X-ray instruments cover different effective energy ranges, we folded the data covering the similar observational time with the same ephemeris and examined the structure change of pulse profiles.
Fig.~\ref{fig:cmp_FLC} shows examples of the comparison between the profiles obtained from different instruments.
There is no obvious phase lag of the major peak determined in soft and hard X-rays, we therefore included all the X-ray data in the TOA analysis.
Each TOA was determined with enough say, $> 300$ counts to resolve the position of the main peak in the pulse profile.
To avoid the unbalanced weighting caused by lots of TOAs determined from {\it XMM} and {\it NuSTAR} over a short time interval, we determined one TOA with the duration of $\sim$5-8\,ks for {\it XMM} data and $\sim$18-20\,ks for {\it Swift} and {\it NuSTAR} observations.

\begin{table}[tp]
\begin{center}
\caption[]{Local ephemeris of \psr\ derived with TOA analysis through X-ray data in Table~\ref{spin}.\\}\label{ephemeris}
\begin{tabular}{ll}
\hline\hline
\multicolumn{2}{l}{Parameter} \\
\hline
Pulsar name\dotfill & \psr\ \\
Valid MJD range\dotfill & 57598--57616 \\
Right ascension, $\alpha$\dotfill & 11:19:14.26* \\
Declination, $\delta$\dotfill &  $-$61:27:49.3* \\
Pulse frequency, $\nu$ (s$^{-1}$)\dotfill & 2.4398374(1) \\
First derivative of pulse frequency, $\dot{\nu}$ (s$^{-2}$)\dotfill & $-2.58(4)\times10^{-11}$ \\
Second derivative of pulse frequency, $\ddot{\nu}$ (s$^{-3}$)\dotfill & $-9(10)\times10^{-19}$ \\
Third derivative of pulse frequency, $\dddot{\nu}$ (s$^{-4}$)\dotfill & $-1.3(1)\times10^{-23}$ \\
Epoch of frequency determination (MJD)\dotfill & 57600 \\
Time system \dotfill & TDB \\
RMS timing residual (ms)\dotfill & 3.5 \\
$\chi_{\nu}^2$/dof\dotfill & 1.30/34 \\
\hline
\hline
Valid MJD range\dotfill & 57598--57632 \\
Pulse frequency, $\nu$ (s$^{-1}$)\dotfill & 2.4398376(2) \\
First derivative of pulse frequency, $\dot{\nu}$ (s$^{-2}$)\dotfill & $-2.5(2)\times10^{-11}$ \\
Second derivative of pulse frequency, $\ddot{\nu}$ (s$^{-3}$)\dotfill & $-1.5(8)\times10^{-18}$ \\
Third derivative of pulse frequency, $\dddot{\nu}$ (s$^{-4}$)\dotfill & $9(3)\times10^{-23}$ \\
Fourth derivative of pulse frequency, $\nu^{(4)}$ (s$^{-5}$)\dotfill & $-3.9(8)\times10^{-28}$ \\
Fifth derivative of pulse frequency, $\nu^{(5)}$ (s$^{-6}$)\dotfill & $8(1)\times10^{-34}$ \\
Sixth derivative of pulse frequency, $\nu^{(6)}$ (s$^{-7}$)\dotfill & $-7.4(8)\times10^{-40}$ \\
RMS timing residual (ms)\dotfill & 3.5 \\
$\chi_{\nu}^2$/dof\dotfill & 1.41/45 \\
\hline
\end{tabular}
\begin{small}
\begin{flushleft}
{\footnotesize
*A FWHM value in 2.3-10 keV is $0.9$\arcsec\ \citep{SK2008}.\\
Note: the numbers in parentheses denote errors in the last digit.\\
More high-order terms are required in an extensive solution to MJD 57632 to describe the timing noise.
}
\end{flushleft}
\end{small}
\end{center}
\end{table} 

From MJD $\sim 57,598$ to MJD~57,616, two high-order polynomial terms ($\ddot{\nu}$ and $\dddot{\nu}$) are required to describe the timing noise, and the result is shown in Table~\ref{ephemeris}.
We note that the derived spin frequency and the spin-down rate are consistent with those determined by \citet{Archibald2016}.
If we extend the effective time range of the pulsar ephemeris for more than half a month, the modified solution can be referred to the lower panel of Table~\ref{ephemeris} and the timing residuals are displayed in Fig.~\ref{fig:residual}. 
We found that more high-order terms are needed, and such a feature can also be applied to characterize the change of the spin-down rate in different time interval as shown in Fig.~\ref{fig:evo_spin}. 
Since the next positive detection after MJD $\sim 57,630$ is one month later and the uncertainties in the timing solution are large, these will lead to cycle count ambiguity in the following TOA analysis and we could not further extend the ephemeris of \psr. 

\subsection{{\rm $\gamma$-Ray Light Curve and Spectrum}}
\label{ssec:Gspectrum}  

\begin{figure}[tp]
  \includegraphics[angle=0,scale=0.36]{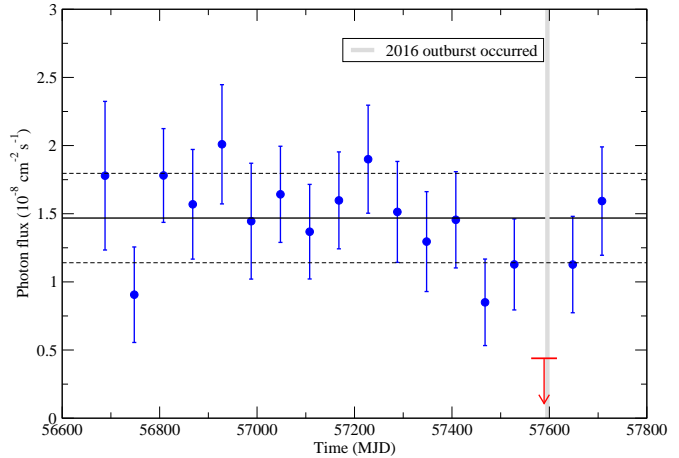} 
\caption{\footnotesize{Variation of the $\gamma$-ray photon flux of \psr. 
Each data point represents the source flux above 500~MeV assessed from a 60-day accumulation of {\it Fermi} observation. 
The black solid and dashed line denote the average flux and 1$\sigma$ uncertainty of the pulsar, respectively.
The source cannot be significantly detected by the data over MJD 57,566--57,626 and the red arrow represents the 3$\sigma$ flux upper limit.
}
\label{fig:Gflux_60day}
}
 \end{figure}
\begin{figure*}[htp]
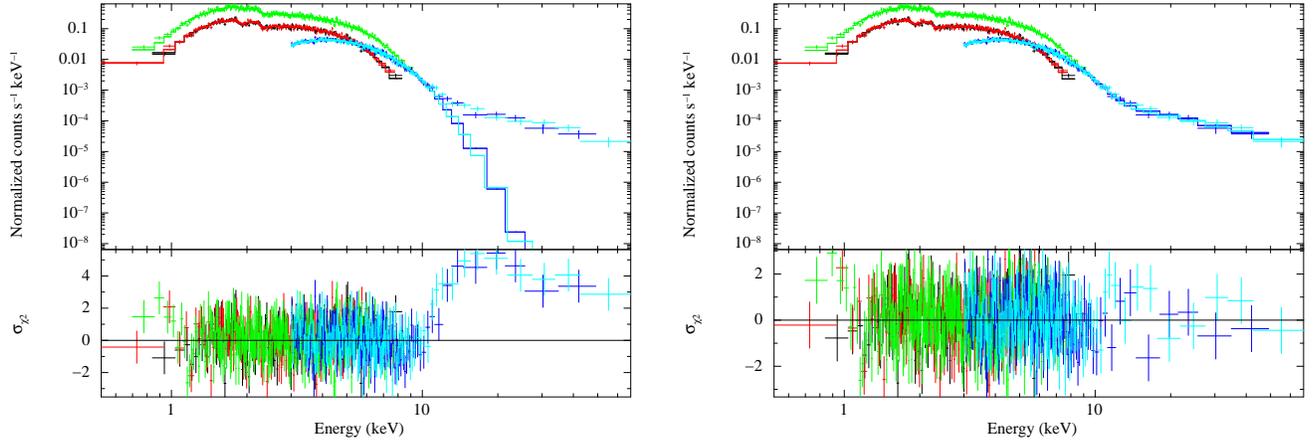

\centering
\hspace*{\fill}{\includegraphics[angle=-90,scale=0.33]{2BB.eps}}
\hspace*{\fill}{\includegraphics[angle=-90,scale=0.33]{2BB+PO.eps}}
\hspace*{\fill}
\caption{\footnotesize Joint fits to the spectra obtained from {\it XMM-Newton}/MOS1 (black), {\it XMM-Newton}/MOS2 (red), {\it XMM-Newton}/PN (green color), {\it NuSTAR}/FPMA (blue) and {\it NuSTAR}/FPMB (cyan) on 2016 Aug 14-15. The left panel displays the fit to the composite model of double blackbody (2BB) components. The right panel presents the fit to the same model with an additional power-law component (2BB+PL). The bottom panel shows the residuals in terms of $\chi^2$.
}
\label{fig:cmp_sp}
\end{figure*}

We performed unbinned likelihood analysis using the ``NewMinunit'' optimization algorithm to derive the flux of our target. 
Spectral parameters of other sources in the region of interest (ROI) is based on the LAT 4-year point source (3FGL; \citealt{3FGL}) catalog, and the standard templates of Galactic and isotropic background (gll\_iem\_v06.fits and iso\_P8R2\_SOURCE\_V6.txt) were also included in our analysis.
The spectral parameters of each source including \psr\ in ROI are thawed, and we assumed the PLEC model for our target.  
We divided the data into 60-day bins to investigate the variation of the source flux.
This result is shown in Fig.~\ref{fig:Gflux_60day}.
Between MJD 57,570--57,630, the pulsar experienced a magnetar-like outburst, but the $\gamma$-ray flux was below the detection level, therefore we only report the flux upper limit.
Even though we considered a smaller time bin (e.g., 30 days) around the outburst or included lower energy band to investigate the $\gamma$-ray flux of \psr, the source flux is still significantly lower than its long-term average. 

The $\gamma$-ray spectrum above 0.5~GeV from MJD 57,560--57,630 can be described by a PLEC model with a photon index of 1.5 and a cutoff energy of 3.1 GeV.
The test-statistic (TS) value \citep{Mattox96} was only 8.4, corresponding to a confidence level less than 3$\sigma$, and it is much less significant than the detection at other epochs.
If we consider a longer accumulation time (i.e., MJD 56,900--57,560) in the pre-outburst stage, the PLEC model has a consistent $\Gamma=1.7\pm0.2$ and a cutoff energy $3.1\pm0.9$ GeV (1$\sigma$ errors).
It seems that the magnetar-like outburst suppressed the $\gamma$-ray emission from the pulsar but the major $\gamma$-ray emission mechanism did not have a significant change.  

Because the post-outburst ephemeris of \psr\ was seriously contaminated by the timing noise, it is difficult to investigate the $\gamma$-ray pulsation through a blind search. 
After the outburst, the effective time range of the pulsar ephemeris determined in the X-ray band (Table~\ref{ephemeris}) is still too short, and {\it Fermi}-LAT cannot accumulate enough photons to detect the $\gamma$-ray pulsation.      
  
\subsection{{\rm Phase-Averaged X-Ray Spectral Analysis}}
\label{ssec:AXspectrum}  
We concentrate the post-outburst spectral analysis with {\it XMM-Newton} and {\it NuSTAR} observations because {\it Swift} XRT and {\it XMM-Newton} EPIC cover a similar energy range but the effective area of {\it Swift}/XRT is much smaller.
The spectrum obtained from each observation was rebinned with different counts (e.g., 25, 50, 75 or 100) per channel to ensure the $\chi^2$ statistic, and the choice of the minimum was decided according to the total photons received from our target.

We fit the {\it XMM-Newton} and {\it NuSTAR} spectra jointly, since they cover a similar observational time interval.
We inspected the spectral behavior of \psr\ within 0.5--65 keV because the number of photons is insufficient beyond this energy range.
We also introduced a constant in the fit to account for the cross-calibration mismatch between three EPIC cameras and two FPMs and applied a photo-electric absorption using Wisconsin cross-sections \citep{MM1983}. 
A single-component model did not give an acceptable result. 
We therefore tried the power-law plus blackbody radiation (BB) model and the 2BB model. 
However, except for the spectra obtained on 2016 Dec. 12--13, the two-component models also do not provide acceptable fits.
We notice that double BB components can provide a good fit to the spectra below 10~keV as shown in the left panel of Fig.~\ref{fig:cmp_sp}, and the excess in the hard X-ray band ($> 10$~keV) can be described with an additional hard power-law.
Even though PL+BB or 2BB model can provide an acceptable fit to the spectra obtained in 2016 mid-Dec, such an excess in the hard X-ray band is still clearly seen.


\begin{table}[bp]
\caption{\small{Best--fit spectral parameters for \psr\ determined from {\it XMM-Newton} and {\it NuSTAR} data.}}\label{ASresult} 
\begin{tabular}{clcccc} 
\hline
\multicolumn{2}{c}{Observed Time} & Aug 05-06 & Aug 14-15 & Aug 30-31 & Dec 12-13 
\\
\hline
 & $N_{\rm H}$$^{a}$  & \multicolumn{4}{c}{1.45$^{+0.07}_{-0.06}$} 
\\ 
 & $\Gamma$ & 0.5$\pm 0.2$ & 0.2$^{+0.3}_{-0.2}$ & 0.5$\pm 0.2$ & -0.6$^{+1.2}_{-1.4}$ 
\\
PL & $F_{\rm{PL}}$$^{b}$  & 0.023 & 0.010 & 0.020 & 0.001 
\\
+ & $kT_1$ (keV) & 0.33$\pm 0.02$  & 0.34$^{+0.01}_{-0.02}$ & 0.33$\pm 0.02$ & 0.35$^{+0.02}_{-0.03}$ 
\\
 $\rm{BB_1}$ & $R_1$ (km)& 4.4$^{+0.8}_{-0.7}$  & 4.0$^{+0.7}_{-0.6}$ & 3.6$^{+0.7}_{-0.6}$ & 1.7$^{+0.4}_{-0.3}$ 
\\
+ & $F_{\rm{BB_1}}$$^{b}$  & 0.34 & 0.28 & 0.21 & 0.058 
\\
 $\rm{BB_2}$ & $kT_2$ (keV) & 1.03$\pm 0.01$ & 1.04$\pm 0.01$ & 1.03$\pm 0.01$ & 1.09$\pm 0.04$
\\
  & $R_2$ (km) & 0.98$\pm 0.02$ & 0.81$\pm 0.02$ & 0.67$\pm 0.02$ & 0.21$\pm 0.02$
\\
 & $F_{\rm{BB_2}}$$^{b}$ & 1.60 & 1.16 & 0.75 & 0.091 
\\
 & $\chi^2_{\nu}$/d.o.f.  & \multicolumn{4}{c}{1.13/2905}
\\
\hline
 & $N_{\rm H}$$^{a}$  & \multicolumn{4}{c}{1.52$\pm 0.03$} 
\\ 
PL & $\Gamma$ & 0.5$\pm 0.2$ & 0.2$^{+0.2}_{-0.3}$ & 0.5$^{+0.2}_{-0.3}$ & 0.2$^{+1.8}_{-1.1}$ 
\\
+ & $F_{\rm{PL}}$$^{b}$  & 0.022 & 0.010 & 0.019 & 0.003 
\\
 NSA$^{c}$ & $kT_1$ (eV) & 234$^{+4}_{-3}$  & 224$^{+4}_{-3}$ & 210$\pm 3$ & 162$^{+2}_{-6}$ 
\\
+ & $F_{\rm{NSA}}$$^{b}$  & 0.44 & 0.36 & 0.28 & 0.086 
\\
 BB & $kT_2$ (keV) & 1.03$\pm 0.01$ & 1.04$\pm 0.01$ & 1.03$\pm 0.01$ & 1.03$^{+0.03}_{-0.04}$
\\
  & $R$ (km) & 0.97$^{+0.01}_{-0.02}$ & 0.81$^{+0.01}_{-0.02}$ & 0.67$^{+0.01}_{-0.02}$ & 0.24$^{+0.01}_{-0.02}$
\\
 & $F_{\rm{BB}}$$^{b}$ & 1.59 & 1.16 & 0.74 & 0.096 
\\
 & $\chi^2_{\nu}$/d.o.f. & \multicolumn{4}{c}{1.14/2909}
\\
\hline
 & $N_{\rm H}$$^{a}$  & \multicolumn{4}{c}{1.55$^{+0.04}_{-0.02}$ } 
\\ 
PL & $\Gamma$ & 0.5$^{+0.2}_{-0.3}$ & 0.1$^{+0.3}_{-0.2}$ & 0.4$^{+0.3}_{-0.2}$  & 0.2$^{+2.8}_{-1.2}$ 
\\
+ & $F_{\rm{PL}}$$^{b}$  & 0.021 & 0.0095 & 0.018 & 0.0030 
\\
 nsmax$^{d}$ & $kT_1$ (eV) & 250$\pm 4$ & 240$^{+3}_{-4}$ & 224$^{+4}_{-3}$ & 172$^{+3}_{-13}$
\\
+ & $F_{\rm{nsmax}}$$^{b}$  & 0.49 & 0.41 & 0.31 & 0.095
\\
 BB & $kT_2$ (keV) & 1.03$\pm 0.01$ & 1.05$\pm 0.01$ & 1.03$\pm 0.01$ & 1.04$^{+0.03}_{-0.05}$
\\
  & $R_2$ (km) & 0.96$^{+0.01}_{-0.02}$ & 0.80$^{+0.01}_{-0.02}$ & 0.66$^{+0.01}_{-0.02}$ & 0.23$^{+0.02}_{-0.01}$
\\
 & $F_{\rm{BB_2}}$$^{b}$ & 1.57 & 1.15 & 0.74 & 0.095 
\\
 & $\chi^2_{\nu}$/d.o.f. & \multicolumn{4}{c}{1.14/2909}
\\
\hline
\end{tabular}
\begin{small}
\begin{flushleft}  
{\footnotesize
$^{a}$ The linked absorption column density is in units of $10^{22}$ cm$^{-2}$.\\
$^{b}$ The unabsorbed flux is measured in 0.5--10~keV and recorded in units of $10^{-11}$~erg~cm$^{-2}$~s$^{-1}$.\\
$^{c}$ We assumed an NS mass of 1.4\,$M_{\odot}$, a radius of 13\,km and the magnetic field of $10^{13}$\,G.
 \\
$^{d}$ 1.4 M$_{\odot}$ in mass and 10~km of radius in size is used to derive the gravitational redshift. The model number is fixed as 1300 for magnetar-like pulsars. The emitting size is the same as c). \\ 
Note: the evolution of each parameter in the PL+2BB model is shown in Fig.~\ref{fig:Jointfit_PO+2BB}.
}
\end{flushleft}
\end{small}
\end{table}

The mid-Dec spectra can be described by a PL+BB model ($\chi_{\nu}^2 = 1.05$ for 283 dof) with a absorption column density ($N_{\rm H}$) of $1.9^{+0.4}_{-0.3}\times 10^{22}$~cm$^{-2}$, a photon index of $3.4^{+0.8}_{-0.5}$ and an emission site of $0.16^{+0.02}_{-0.01}$~km in radius with a blackbody temperature of $1.18^{+0.04}_{-0.06}$~keV.
All the quoted errors of X-ray spectral parameters in this article correspond to 90\% confidence interval for one parameter of interest, and a source distance 8.4\,kpc was assumed \citep{CMC2004}.  
The measured photon index is softer than the {\it Chandra} spectrum obtained in 2016 late Oct \citep{BSM2017}.
The unabsorbed 0.5--7\,keV flux derived from our spectral model is $(3.3 \pm 1.8)\times 10^{-12}$~erg~cm$^{-2}$~s$^{-1}$, which is about 30\%--50\% lower than the non-thermal flux derived from the subsequent {\it Chandra} observation.
The non-thermal component can be originated from the magnetospheric emission.
On the other hand, we note that the soft X-ray photons (i.e., $< 2.5$~keV) of this component contribute to a substantial fraction of the pulsed photons to lead to a large pulsed fraction ($\gtrsim 50$\%), indicating a non-thermal pulse detection.
However, the pulse profiles determined in the soft and medium X-ray (i.e., 2.5--10 keV; dominated by BB component) bands have the same shape, which does not support a strong non-thermal component existed in the spectra.  

\begin{figure*}[htp]
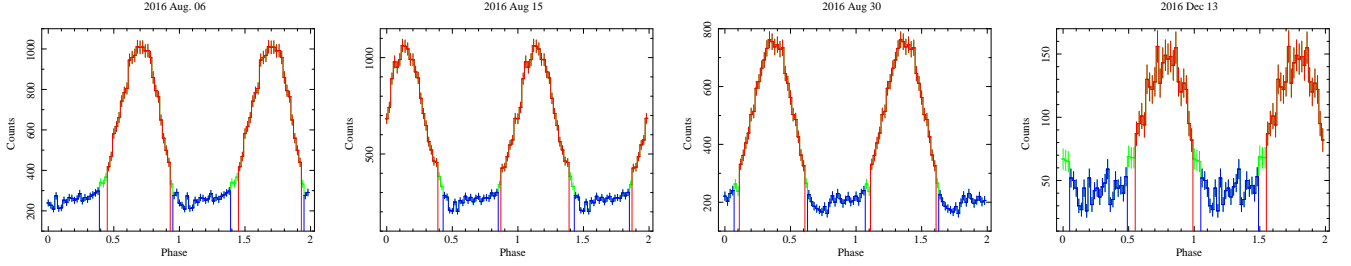

\centering
\hspace*{\fill}{\includegraphics[scale=0.17,angle=-90]{PF_comp1.eps}}
\hspace*{\fill}{\includegraphics[scale=0.17,angle=-90]{PF_comp2.eps}}
\hspace*{\fill}{\includegraphics[scale=0.17,angle=-90]{PF_comp3.eps}}
\hspace*{\fill}{\includegraphics[scale=0.17,angle=-90]{PF_comp4.eps}}
\hspace*{\fill}
\caption{\small Folded light curves of \psr\ in 0.15--12~keV obtained from four {\it XMM-Newton}/PN observations. It is folded light curve at at MJDs 57606.6, 57616.0, 57630.3 and 57735.5, with spin frequency of 2.4398214\,Hz, 2.4397945\,Hz, 2.4397218\,Hz and 2.4391766\,Hz, respectively. The region under the red and blue curves represent the on- and  off-pulse phases, respectively, in each profile. Two cycles of each profile are shown for clarity.}
\label{fig:xmmFLC}
\end{figure*}

Different from the post-outburst spectral fits to {\it NuSTAR} and {\it Swift}/XRT data of late July, we cannot obtain an acceptable fit with PL+BB model except for the spectra obtained in mid-Dec. 
We tried fitting the PL+BB model to the Dec 12--13 spectra, but obtained a totally different photon index ($3.4^{+0.8}_{-0.5}$) comparing with that on July 28--29 ($1.2\pm 0.2$) determined in \citet{Archibald2016}. 
The non-thermal component in our spectra dominates the soft X-ray photons, but it in the fit to the spectra of {\it NuSTAR} and {\it Swift}/XRT mainly describes the hard X-ray tail.
Because the PL+BB model also cannot provide acceptable fits to the post-outburst spectra obtained in 2016 Aug, we rejected it in our investigations.
Table~\ref{ASresult} lists the best-fit parameters of the three-component model, and this model is rather common to describe the spectral behavior for magnetars \citep{MN2018}.
Fig.~\ref{fig:cmp_sp} shows the fit to the mid-Aug spectra, and it clearly indicates that a composite model of two thermal and one power-law components (2BB+PL) can well describe the observational features.
If we only consider the {\it NuSTAR} spectra in 3--65~keV, the composite model with a hard PL and a single BB component can give a good fit, and it may indicate that the cooler blackbody component was only detected by {\it XMM-Newton} due to its high sensitivity at low energy.    

The double BB components determined in the 2BB+PL model can be characterized by a smaller emission site (i.e. $\lesssim$ 1~km in radius) with a higher temperature of $\sim$~1.0~keV and a larger emission site of a few kilometers in radius with a lower temperature of 0.3--0.4 keV.
The best-fit interstellar absorption determined with three-component model is $1.45^{+0.07}_{-0.06}\times 10^{22}$~cm$^{-2}$, and this value is between the post-outburst absorptions determined by \citet{Archibald2016} and \citet{BSM2017}.
Comparing with the pre-outburst absorption, this value is also consistent with that from {\it Chandra} observations \citep{SK2008} but obviously lower than that determined by the joint fit to both {\it XMM-Newton} and {\it Chandra} spectra \citep{Ng2012}.       
We find that the emission sizes of two hotspots can be larger but the hotspot temperature is similar if the absorption column density is fixed at a larger value during the fit.    
 
In addition to the blackbody emission, we also considered NS atmosphere models, with a pure hydrogen atmosphere (NSA; \citealt{ZPSV95}) and an atmosphere with hydrogen or other heavy elements (nsmax; \citealt{HPC2008}).
\psr\ has strong magnetic field of $\sim 4.1\times 10^{13}$~G, therefore we adopted a field strength of the same order (i.e., $10^{13}$~G) in both the NSA and the nsmax models. 
Comparing with the two BB components in the three-component model, only the lower temperature thermal component can be replaced by the emission from the magnetized atmosphere.        
If we considered the change of $N_{\rm H}$ at different epochs for spectral fits to the PL+NSA+BB and the PL+nsmax+BB model, the best fits to both models give a consistent value of $\sim$~(1.4--1.6) $\times 10^{22}$~cm$^{-2}$ in Aug but a relatively higher value of $\sim$~(1.8--2.0) $\times 10^{22}$~cm$^{-2}$ in mid-Dec.  

We did not find any obvious decreasing trend of the temperature as presented by \citet{Archibald2016} among the spectral fits to a composite model with double BB components shown in Table~\ref{ASresult}.
We detected a trend with both hotspots shrank and their flux contribution decreased after the magnetar-like outburst.
The evolution of the flux ratio between the hotter and the cooler components could indicate the gradual disappearance of the hotter spot.
The decrease of the temperature can be detected if we replaced the cooler thermal component by the global atmospheric emission with a uniform temperature.                   
         
\subsection{{\rm Phase-Resolved X-Ray Spectral Analysis}}
\label{ssec:PrXspectrum} 

Based on the result of the phase-averaged spectral analysis shown in Table~\ref{ASresult}, we find that the two thermal components dominate the spectral behavior in the energy range $\lesssim 10$~keV. 
The background subtracted pulsed fractions can be determined by $(f_{max} - f_{min})/(f_{max} + f_{min})$, where $f_{max}$ and $f_{min}$ are the maximum and minimum counts in folded bins, respectively.
These values inferred from four {\it XMM-Newton}/PN data are $66\pm 2$\%, $70\pm 2$\%, $67\pm 3$\% and $87\pm 5$\% while the {\it NuSTAR} observations give $65\pm 2$\%, $61\pm 2$\%, $49\pm 3$\% and $74\pm 11$\% at similar epochs.
The pulsed fraction obtained from the {\it XMM-Newton}/PN observations is higher and no pulsation can be detected $>10$~keV after end of July, thus, we speculate that the additional PL component required in the joint spectral analysis with {\it NuSTAR} observations has no contribution to the pulsations of \psr.    
Accordingly, we only generated the pulsed spectra of different epochs from the {\it XMM-Newton}/PN observations.

As shown in Fig.~\ref{fig:xmmFLC}, we defined the phase intervals of 0.46--0.92, 0.88--1.38, 0.12--0.60 and 0.56--0.98 as the ``on-pulse'' components for data observed on Aug 6, Aug 15, Aug 30 and Dec 13 of 2016, respectively. 
The ``off-pulse'' (i.e., DC level) interval in the folded light curve was then determined at the phase within 0.96--1.38, 0.44--0.84, 0.64--1.06 and 0.06--0.48 for data observed on Aug 6, Aug 15, Aug 30 and Dec 13 of 2016, respectively.
We generated pulsed spectra at different epochs by subtracting the ``off-pulse'' spectra from the ``on-pulse'' spectra.
In order to ensure the $\chi^2$ statistic, we re-grouped the channels to have at least 50, 80, 50 and 25 photons per channel for each observation gained from 2016 early Aug to mid-Dec.
We also generated the unpulsed spectra by subtracting the background contribution from the source spectrum gained within the ``off-pulse'' interval as defined in Fig.~\ref{fig:xmmFLC}, and the background spectrum was determined by the nearby source free region.
The unpulsed spectrum was also rebinned with a minimum of 25, 40, 25 and 15 counts per channel for each dataset observed from 2016 early Aug to mid-Dec.


\begin{table}[bp]
\caption{\small{Best-fit parameters for the pulsed spectra of \psr\ obtained from {\it XMM-Newton}/PN.}}\label{PSresult} 
\begin{tabular}{clcccc} 
\hline
\multicolumn{2}{c}{Observed Time} & Aug 06 & Aug 15 & Aug 30 & Dec 13 
\\
\hline
 & $N_{\rm H}$$^{a}$  & \multicolumn{4}{c}{0.85$\pm 0.04$} 
\\
BB & $kT$ (keV) & 1.01$\pm 0.02$ & 0.96$\pm 0.02$ & 0.93$^{+0.02}_{-0.03}$ & 0.79$\pm 0.05$ 
\\
 & $R$ (km) & 1.01$\pm 0.04$ & 0.90$^{+0.04}_{-0.03}$ & 0.78$^{+0.03}_{-0.04}$ & 0.40$\pm 0.04$
\\
 & $F_{\rm{BB}}$$^{b}$ & 1.52 & 1.02 & 0.66 & 0.093
\\
  & $\chi^2_{\nu}$/d.o.f. & \multicolumn{4}{c}{1.25/569}  
\\
\hline
 & $N_{\rm H}$$^{a}$  & \multicolumn{4}{c}{1.8$\pm 0.2$} 
\\
PL & $\Gamma$ & 3.2$^{+0.7}_{-0.5}$ & 3.0$\pm 0.5$ & 2.7$^{+0.6}_{-0.4}$ & 3.2$^{+0.7}_{-0.6}$ 
\\
+ & $F_{\rm{PL}}$$^{b}$  &1.58 & 1.19 & 0.93 & 0.27
\\
BB & $kT$ (keV) & 1.00$\pm 0.03$ & 0.97$^{+0.03}_{-0.04}$  & 0.96$^{+0.07}_{-0.09}$ & 1.2$^{+0.2}_{-0.2}$ 
\\
 & $R$ (km) & 0.99$^{+0.09}_{-0.08}$& 0.81$^{+0.07}_{-0.08}$ & 0.58$^{+0.1}_{-0.09}$ & 0.13$^{+0.06}_{-0.07}$  
\\
 & $F_{\rm{BB}}$$^{b}$ & 1.42 & 0.84 & 0.41 & 0.044
\\
  & $\chi^2_{\nu}$/d.o.f. & \multicolumn{4}{c}{1.10/561} 
\\
\hline
 & $N_{\rm H}$$^{a}$  &  \multicolumn{4}{c}{1.3$^{+0.1}_{-0.2}$ }  
\\ 
 & $kT_1$ (keV) & 0.37$^{+0.12}_{-0.07}$ & 0.45$^{+0.10}_{-0.08}$ & 0.46$^{+0.11}_{-0.08}$ & 0.38$^{+0.06}_{-0.05}$ 
\\
 $\rm{BB_1}$ & $R_1$ (km)& 3$^{+2}_{-1}$  & 2.1$^{+1.0}_{-0.7}$ & 1.9$^{+0.8}_{-0.6}$ & 1.4$^{+0.6}_{-0.4}$ 
\\
 + & $F_{\rm{BB_1}}$$^{b}$  & 0.24 & 0.25 & 0.24 & 0.060 
\\
 $\rm{BB_2}$ & $kT_2$ (keV) & 1.01$^{+0.05}_{-0.03}$ & 1.03$^{+0.06}_{-0.05}$ & 1.08$^{+0.11}_{-0.08}$ & 1.2$^{+0.1}_{-0.2}$
\\
  & $R_2$ (km) & 1.00$^{+0.08}_{-0.10}$ & 0.76$^{+0.09}_{-0.11}$ & 0.5$\pm 0.1$ & 0.17$\pm 0.05$
\\
 & $F_{\rm{BB_2}}$$^{b}$ & 1.53 & 0.94 & 0.55 & 0.071 
\\
 & $\chi^2_{\nu}$/d.o.f. & \multicolumn{4}{c}{1.07/561} 
\\
\hline
\end{tabular}
\begin{small}
\begin{flushleft}
{\footnotesize
$^{a}$ The linked absorption column density is in units of $10^{22}$ cm$^{-2}$.\\
$^{b}$ The unabsorbed flux is measured in 0.5--8~keV and recorded in units of $10^{-11}$~erg~cm$^{-2}$~s$^{-1}$.\\
}
\end{flushleft}
\end{small}
\end{table}

\subsubsection{{\rm Pulsed Spectral Analysis}}

A single PL or a single BB model does not provide a good fit to most of the individual pulsed spectrum.  
Although the PL+BB model can provide acceptable fits to the pulsed spectra, the photon index is abnormally large (i.e., $> 2.5$) comparing to the general non-thermal X-ray emission from pulsars \citep{CZ99}. 
In addition, the non-thermal component both dominates the source flux and the spectra in the soft X-ray band.
We infer that the dominant non-thermal component also contributes to the pulsed emission since it was obtained from the pulsed spectra. 
The non-thermal pulsations corresponding to a different physical origin should have a quite distinct profile than the thermal pulsation originated from the surface emission.
For example, the synchrotron radiation in the magnetosphere have a pulse profile similar to what we detected in the hard X-ray band with {\it NuSTAR} (Fig.~\ref{fig:ERFLC}), and it is different from the profile obtained from BB component.
Nevertheless, we obtain similar structure without any phase change in the energy-resolved pulse profiles below 10~keV, and it disfavors the existence of different physical origins in the pulsed spectra.

We present the simultaneous fit to the pulsed spectrum with 2BB components and a linked hydrogen absorption in Table~\ref{PSresult}.
The thermal flux derived for the two hotspots and the inferred emission size are somewhat lower/smaller than what we obtain in Table~\ref{ASresult}, but are still statistically consistent.
Because the composite model contributed by two thermal components provides a better fit and a more physical origin, this is our preferred model.
We cannot conclude a significant change in the temperature of each thermal component derived from different observations, but the shrinking of both emitting regions led to a decreasing pulsed emission with time.


\begin{table}[bp]
\caption{\small{Best-fit parameters for the unpulsed spectra of \psr\ obtained from {\it XMM-Newton}/pn.}}\label{UPSresult} 
\begin{tabular}{clcccc} 
\hline
\multicolumn{2}{c}{Observed Time} & Aug. 06 & Aug. 15 & Aug. 30 & Dec. 13 
\\
\hline
 & $N_{\rm H}$$^{a}$  & \multicolumn{4}{c}{0.79$^{+0.04}_{-0.05}$} 
\\
BB & $kT$ (keV) & 1.09$^{+0.03}_{-0.02}$ & 1.11$\pm 0.02$ & 1.11$\pm 0.03$ & 1.1$\pm 0.1$
\\
 & $R$ (km) & 0.61$^{+0.02}_{-0.03}$ & 0.51$\pm 0.02$ & 0.41$\pm 0.02$ & 0.12$\pm 0.02$
\\
 & $F_{\rm{BB}}$$^{b}$ & 0.74 & 0.55 & 0.36 & 0.033 
\\
  & $\chi^2_{\nu}$/d.o.f. & \multicolumn{4}{c}{1.09/488}   
\\
\hline
 & $N_{\rm H}$$^{a}$  & \multicolumn{4}{c}{1.7$\pm 0.3$} 
\\
PL & $\Gamma$ & 2.2$\pm 0.5$ & 3.0$^{+0.8}_{-0.7}$ & 2.9$\pm 0.8$ & 3$^{+5}_{-2}$ 
\\
+ & $F_{\rm{PL}}$$^{b}$  & 0.66 & 0.46 & 0.28 & 0.035
\\
BB & $kT$ (keV) & 1.05$^{+0.09}_{-0.10}$ & 1.13$^{+0.04}_{-0.03}$ & 1.12$^{+0.04}_{-0.05}$ & 1.2$\pm 0.5$ 
\\
 & $R$ (km) & 0.53$^{+0.10}_{-0.08}$ & 0.46$\pm 0.05$ & 0.38$\pm 0.04$ & 0.09$^{+0.06}_{-0.07}$
\\
 & $F_{\rm{BB}}$$^{b}$ & 0.48 & 0.49 & 0.31 & 0.023
\\
  & $\chi^2_{\nu}$/d.o.f. & \multicolumn{4}{c}{1.01/480} 
\\
\hline
 & $N_{\rm H}$$^{a}$  & \multicolumn{4}{c}{1.2$\pm 0.2$} 
\\ 
 & $kT_1$ (keV) & 0.6$^{+0.2}_{-0.1}$ & 0.37$^{+0.11}_{-0.07}$ & 0.40$^{+0.15}_{-0.09}$ & 0.4$^{+0.6}_{-0.2}$ 
\\
 $\rm{BB_1}$ & $R_1$ (km)& 1.2$^{+0.6}_{-0.4}$  & 1.8$^{+1.4}_{-0.8}$ & 1.2$^{+0.5}_{-1.0}$ & 0.5$^{+2.3}_{-0.4}$ 
\\
 + & $F_{\rm{BB_1}}$$^{b}$  & 0.25 & 0.083 & 0.053 & 0.0082
\\
 $\rm{BB_2}$ & $kT_2$ (keV) & 1.3$^{+0.3}_{-0.1}$ & 1.14$^{+0.05}_{-0.04}$& 1.14$^{+0.06}_{-0.05}$ & 1.23$^{+0.04}_{-0.18}$
\\
  & $R_2$ (km) & 0.4$^{+0.1}_{-0.2}$ & 0.48$^{+0.04}_{-0.05}$ & 0.39$^{+0.03}_{-0.05}$  & 0.10$^{+0.04}_{-0.10}$
\\
 & $F_{\rm{BB_2}}$$^{b}$ & 0.58 & 0.54 & 0.35 & 0.031 
\\
 & $\chi^2_{\nu}$/d.o.f. & \multicolumn{4}{c}{1.00/480}  
\\
\hline
\end{tabular}
\begin{small}
\begin{flushleft}
{\footnotesize
$^{a}$ The linked absorption column density is in units of $10^{22}$ cm$^{-2}$.\\
$^{b}$ The unabsorbed flux is measured in 0.5--8~keV and recorded in units of $10^{-11}$~erg~cm$^{-2}$~s$^{-1}$.\\
Note: the evolution of each parameter in PL+BB and 2BB models can be referred to Figs.~\ref{fig:Unpulsed_PO+BB} and \ref{fig:Unpulsed_2BB}.
}
\end{flushleft}
\end{small}
\end{table}

\subsubsection{{\rm Unpulsed Spectral Analysis}}

The unpulsed spectra of \psr\ extracted from the ``off-pulse'' phase can be adequately fit by a single BB model and the results are listed in Table~\ref{UPSresult} in contrast to the phase-averaged spectra and the pulsed spectra.
However, the fit prefers a small $N_{\rm H}$, lower than all previously reported values (e.g., $(1.1\pm 0.1)\times 10^{22}$~cm$^{-2}$; \citealt{BSM2017}).
Adding another BB component gives a larger absorption (see Table~\ref{UPSresult}). 
If we fix $N_{\rm H}$ at a higher value, we obtain lower temperatures for the BB components, e.g. $N_{\rm H}=1.6\times 10^{22}$~cm$^{-2}$ gives $kT=0.43^{+0.06}_{-0.07}$ and $1.18^{+0.07}_{-0.06}$\,eV for 2016 Aug 6 data.
The thermal component at the off-pulse phase has a similar temperate, but a smaller effective radius (13--70\% for the cooler component and 16--60\% for hotter component) than that at the on-pulse phase.  
Hence, we speculate that only part of thermal emission can be observed at the off-pulse phase through a light bending effect due to the strong gravity, even though the hotspot is not in the line of sight.

BB+PL model can also provide acceptable fits to the spectra.
However, NSA+PL does not provide a good fit to the spectrum.
The best-fit PL+BB model has $N_{\rm H}=(1.7\pm 0.3)\times 10^{22}$~cm$^{-2}$ and a large photon index ($\Gamma \gtrsim 3$). 
Such a large absorption was also obtained in the pre-outburst spectral analysis of \psr. 
The unabsorbed flux contributed by the thermal and non-thermal components are comparable.
We note that this PL component could not be contributed from PWNe because such a large photon index was rarely obtained by the non-thermal X-rays from PWNe \citep{KP2008}.
However, the soft non-thermal component could be originated from the magnetospheric upscattering of the thermal photons \citep{TLK2002}.

\section{Discussion}
\label{sec:discussion}  
The positive detections of X-ray pulsations for \psr\ after 2016 early Sep are few, and we cannot extend the effective time range of the ephemeris (see Table~\ref{ephemeris}) through the TOA analysis.
According to the spin-down rate derived by the linear regression over different time ranges, our results did not show any hints of its recovery.
Instead of the spin-down recovery that can be observed after two previous glitches \citep{WJE2011,Anton2015} or after the glitch event of other high-$B$ pulsar \citep{Livingstone2011}, we suspect that some events occurred at the end of 2016 Aug could have slowed down the pulsar spin and continuously increased its spin-down rate.
Three short magnetar-like X-ray bursts detected on 2016 Aug 30 may relate to not only the change of the radio pulse profile \citep{Archibald2017} but also the increase of the timing noise and the spin-down rate.

\citet{Dai2018} observed the pulsar regularly with the Parkes radio telescope and presented a complete evolution of the timing solution.
According to their results, both the radio flux and the spin-down rate of the pulsar continued to increase after the outburst at the end of July until the end of Aug.
The complicated timing behavior well followed the expectation of our timing analysis although we do not have enough dense data sets to trace the recovery of the spin-down rate starting from the end of Aug.
During the recovery process, a two-component radio pulse profile transformed into a four-component one accompanying a sudden flux drop in the radio band.

We need two thermal components to fit the X-ray spectra in our spectral analysis.
An NS with a strong toroidal magnetic field can lead to the discontinuities in the tangential $B$-field component, where the Ohmic dissipation is strongly enhanced \citep{VP2012}, affecting the thermal evolution through Joule heating.
The heat conduction then becomes anisotropic, resulting in inhomogeneous surface temperature distributions \citep{Vigano2013}.
The thermal pulse profile can eventually evolve into single-peaked with a large pulsed fraction \citep{Perna2013}; however, \psr\ is a young pulsar with a characteristic age of only 1600\,yr.
The cooler surface emission can be hidden under the case of strong interstellar absorption ($N_{\rm H} \gtrsim 10^{22}$ cm$^{-2}$), and the detected thermal component of the NS within the energy band of 0.5--2 keV only reveals on a single hotspot of about 1--2 km (cf. simulations in \citealt{SL2012}).   
Nevertheless, the cooler hotspot obtained in our X-ray spectral fit has a temperature of $> 0.3$~keV and contributes to a significant fraction of photons below 2.5~keV, so a more complete model to re-investigate how the multiple thermal components lead to a single-peaked profile is required.
The contribution of a cooler but larger thermal component can also be described by the global magnetized atmospheric emission with a uniform temperature.
Because the cooler thermal emission also has a substantial contribution to the pulsations, we speculate that the emission scenario from two local hotspots could explain the observed pulsed structure. 
We also note that the hotter thermal emission in the composite model can be replaced by a Comptonization component (i.e., CompBB model; \citealt{NMI86}), and the blackbody temperature of the Comptonized model is consistent to that determined for the surface emission.        

\begin{figure}[tp]
  \includegraphics[angle=0,scale=0.36]{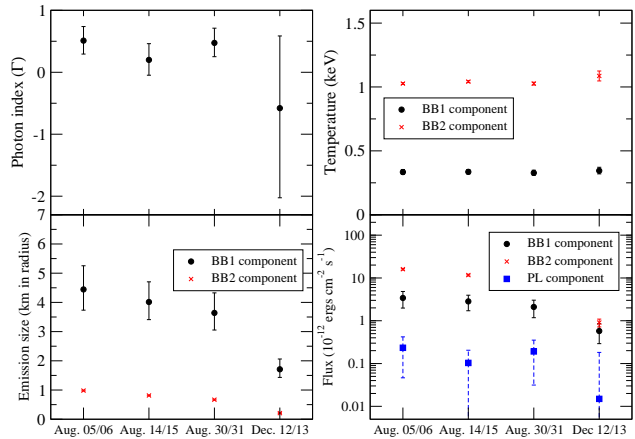} 
\caption{\footnotesize{Spectral parameters obtained by the simultaneous fit to the joint spectra of {\it XMM-Newton} and {\it NuSTAR} observations with the PL+2BB model and a linked absorption. The parameters are listed in Table~\ref{ASresult}. The error bars denote 90\% confidence intervals.    
}
\label{fig:Jointfit_PO+2BB}
}
 \end{figure}
 
We cannot differentiate the contribution of the non-thermal component dominated in the soft X-ray pulsation (i.e., $< 2.5$~keV) from the thermal pulsation although the BB+PL model can also provide an acceptable fit to the phase-averaged spectra below 10~keV.
Comparing with the flux of PWN detected in the post-outburst stage \citep{BSM2017} and the time required to power it by the 2016 bursts, the unabsorbed flux contribution larger than $10^{-12}$~erg~cm$^{-2}$~s$^{-1}$ with a steep photon index of $\Gamma \gtrsim 3$ should result from the different origin, even if we cannot exclude a negligible contribution from the PWN.
The similar combined components can also be resolved in the pulsed spectra extracted from the {\it XMM-Newton} observations as shown in Table~\ref{PSresult}.
However, a non-thermal origin to emit a large amount of soft X-ray photons is not preferred to describe the pulsed emission because we did not find a pulsed structure change comparing with the thermal origin.  
An additional PL component with $\Gamma < 1$ is indeed required to describe the spectral behavior in the hard X-ray band (i.e., $> 10$~keV) when we further include the spectra obtained by {\it NuSTAR} observations.
Because the soft photon index ($2.2\pm 0.5$) determined by the spectrum of PWN in the post-outburst stage \citep{BSM2017} is much softer, we argue that this component should be attributed to the magnetospheric emission of \psr.
This component cannot result from the outer gap emission of the pulsar after the end of 2016 July because 
we detected more than 2000 source counts over 10\,keV but the pulsations were not significant at all.
Such a hard component was also detected in some persistently bright magnetars (e.g., 4U 0142+61) or from the spectral slope changes of some transient magnetars (e.g., 1E 1547.0$-$5408) during outbursts \citep{Enoto2017}. 

\begin{figure}[bp]
  \includegraphics[angle=0,scale=0.36]{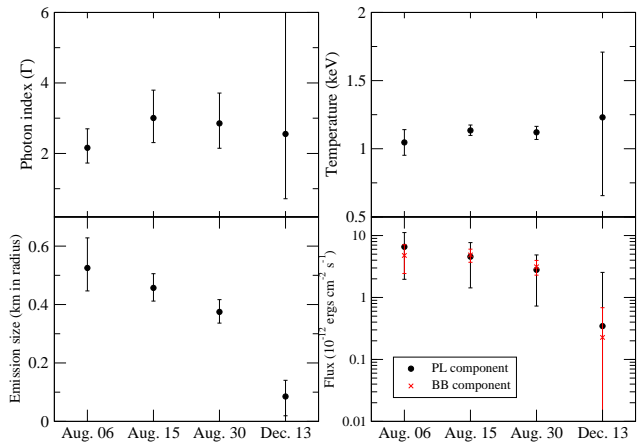} 
\caption{\footnotesize{Spectral parameters obtained by the simultaneous fit to the unpulsed spectra with a PL+BB model components and a linked absorption. 
The parameters are listed in Table~\ref{UPSresult}. The error bars denote 90\% confidence intervals.    
}
\label{fig:Unpulsed_PO+BB}
}
 \end{figure}
\begin{figure}[h]
  \includegraphics[angle=0,scale=0.36]{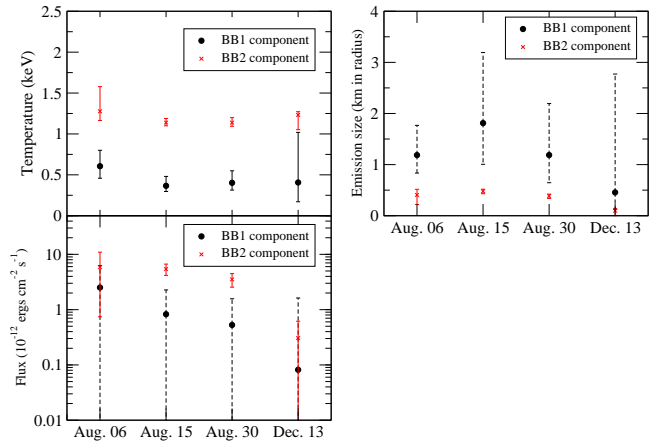} 
\caption{\footnotesize{Spectral parameters obtained by the simultaneous fit to the unpulsed spectra with 2BB components and a linked absorption. 
The parameters are listed in Table~\ref{UPSresult}. The error bars denote 90\% confidence intervals.    
}
\label{fig:Unpulsed_2BB}
}
 \end{figure}

Both of the phase-averaged spectral and the pulsed spectral analyses did not show an obvious temperature variation for the two thermal components, but we can detect a trend of decreasing emission area, as shown in Fig~\ref{fig:Jointfit_PO+2BB}.
A similar feature was also observed in the outburst decay of some magnetars \citep{Esposito2013,Castillo2016}.
The decrease in temperature can only be seen when the cooler BB component with the temperature of $0.3-0.4$~keV is replaced by the NS atmosphere model.
When we perform simultaneous fit to the pulsed spectrum with a linked absorption ($N_{\rm H}=1.59^{+0.07}_{-0.09}\times 10^{22}$~cm$^{-2}$) with the NSA+BB model, we found a uniform temperature over the entire surface of a 1.4~M$_{\odot}$ NS with 13 km radius evolved with $245^{+13}_{-9}$~eV, $234^{+9}_{-10}$~eV, $223^{+8}_{-9}$~eV and $172^{+5}_{-6}$~eV obtained from the {\it XMM-Newton} data on Aug. 06, Aug. 15, Aug. 30 and Dec. 13, respectively.
   
In the unpulsed spectral analysis, both the composite models, PL+BB and 2BB, provide acceptable fits with the spectral parameters shown in Figs.~\ref{fig:Unpulsed_PO+BB} and \ref{fig:Unpulsed_2BB}.
The resonant cyclotron scattering of the surface thermal emission \citep{TLK2002} and the gravitational bending of the emission from the warmer hotspot may provide the radiative scenario of the PL+BB model.
The non-thermal emission, which dominates the soft X-ray photons (i.e., $< 2.5$~keV), is comparable to the surface emission in this model; nevertheless, the effect of resonant Compton scattering mainly produces the hard X-ray photons and is difficult to explain the off-pulse emission with such a soft PL component.
It is also strange that both the pulsed and off-pulsed emission have the similar emission origin, but we note that the emission sizes of two hotspots determined by the unpulsed spectra (i.e., from 1.8 to 0.5\,km for the larger hotspot radius and from 0.48 to 0.10\,km for the smaller one) are smaller that those inferred from the fit to the pulsed spectrum (i.e., from 3.0 to 1.4\,km for the larger hotspot radius and from 1.00 to 0.17\,km for the smaller one). 
This result suggests that the emission from hotspots, which is not in the direction of our sight, can still be partially observed due to the gravitational light bending and contributes weaker thermal components in the spectra during the off-pulse phase.
  
Around the X-ray outburst of \psr, we found the source flux $F_{0.5-300 \rm{GeV}} < 4.9\times 10^{-12}$~erg cm$^{-2}$ s$^{-1}$, which is clearly lower than $(1.2\pm 0.1)\times 10^{-11}$~erg cm$^{-2}$ s$^{-1}$ determined in the pre-outburst stage.
Although the significance of the detection was clearly decreased, similar $\gamma$-ray spectral behavior can be inspected before and during the outburst stages.
It indicates that the acceleration mechanism to emit $\gamma$-ray photons was suppressed without a significant change in the emission geometry.
The theoretical model for the evolution of the $\gamma$-ray and the X-ray emission after the outburst will be fully discussed in our subsequent paper.   
Similar suppression after the outburst can also be detected in the radio band at the end of 2016 July or early Aug \citep{Majid2017}.

It is not clearly understood what caused the increase of the spin-down rate with a month time scale and the disappearance of the radio emission after the glitch, but it could be related to a reconfiguration of the open field line region \citep{Akbal2015,Parfrey2012,HYT2016}.
Changes of the global magnetospheric configuration is also expected for some transient RPPs, for which two emitting states with different spin-down states have been observed.
For example, PSR B1931+24 transferred between radio-bright and radio-quiet modes with a time scale from days to weeks  \citep{Kramer2006}.  
PSR~J2021+4026 showed a mode change in $\gamma$-rays during a glitch in 2011 and had stayed at a low $\gamma$-ray luminosity with high spin-down rate for about three years \citep{Zhao2017}. 
Moreover, PSR B0943+10 showed synchronous mode-switching between the radio and the X-ray emission \citep{Hermsen2013,Mereghetti2016}. 
We note that a similar synchronous change can be detected for \psr\ after the 2016 outburst.
The X-ray background subtracted pulsed fractions of \psr\ obtained from {\it NuSTAR} on July 28-29 and on Aug. 30-31 are $63\pm 2$\% and $49\pm 3$\%, which could also indicate a relatively higher ratio of non-thermal contribution when the radio emission was switched on.  
Although the physical origin of our detections can be different from RPPs and the mode change of the latter could be a state change of the global magnetospheric current, triggers of the state changes in the electromagnetic emission and/or the spin-down rate are not yet well understood and could depend on the type of the transients.

It is difficult to evaluate the effect on the timing solution due to the short magnetar-like bursts occurred at the end of 2016 Aug in our study, so we cannot definitively conclude that the larger spin-down rate was caused by the reconfiguration of the global magnetosphere.
Another interpretation is that a large amount of electron-positron pairs created in the trapped photon-pair plasma fireball cease the entire acceleration mechanism of an NS \citep{TD95}.
Such a scenario was also applied to explain the sudden radio drop over a few tens of seconds coincident with the occurrence of multiple short X-ray bursts detected at the end of 2016 Aug \citep{Archibald2017}.
The series of X-ray bursts occurred at the end of 2016 July can also provide the energy of the trapped fire ball to create the pairs to suppress the radio and high-energy emission. 
The major difference after the X-ray bursts at the end of July and Aug may be the cooling time scale of the pair plasma in the trapped fireball because no significant radio detection was obtained until 2016 Aug. 1 \citep{Majid2017} and it seems that the recovery of the radio emission required a longer time.

\section{Conclusion}
\label{sec:conclusion}
We have performed a complete study to trace the high-energy timing and spectral evolution of \psr\ after its 2016 outburst.
Based on the confirmed periodic signals detected at different epochs, we derived the timing ephemeris of \psr\  after the outburst.
We also clearly found the decreasing trend of the spin frequency derivative after the glitch/outburst, which is different from the recovery of the spin-down rate generally observed following similar events, or even in the previous glitches of 2004 and 2007. 
The serious timing noise embedded in the timing solution can be detected at the end of 2016 Aug by comparing ephemerides determined within different time interval in Table~\ref{ephemeris}, and it also prohibits blind searches of pulsations in $\gamma$-rays.

In the beginning of the 2016 outburst, the X-ray emission of $< 10$~keV in the phase-averaged spectroscopy can be mainly described with a single thermal component.
After the outburst, we found that the spectrum requires two thermal components to describe X-rays from different origins; the global atmosphere/large emitting region (with a few kilometers in radius) and from the small hotspot (with the radius less than one kilometer). 
Comparing with the spectral behavior of the pulsar before the 2016 outburst, the pulsed fraction in $2.5-8$~keV significantly increased and an additional component is required to fit the spectrum.
Instead of a decrease in the temperature after the outburst, we found that the emission sites shrank to $\sim 65$\% and 80\% of their original size in four months.
A shrinking of the thermal emitting region is also detected to be accompanied by a decrease of the flux. 
The decrease of the temperature is only significant when we replace the cooler thermal component of $\sim 0.3-0.4$~keV by the magnetized atmospheric emission, but we did not prefer such an interpretation because the pulsed emission in the soft X-ray band ($< 2.5$~keV) was quite significant, inconsistent with a global emission scenario.
We also need to include a hard PL component ($\Gamma < 1$) to describe the hard X-ray ($>10$~keV) spectrum from {\it NuSTAR}.    
Such a component with an unclear origin cannot be detected in RPPs in general; even another high-$B$ pulsar, PSR~J1846$-$0258 shows an obvious softening of the non-thermal emission in the outburst \citep{NSGH2008}.
Comparing with strong thermal emission from the pulsar, the emission from the PWN and magnetospheric pulsation is low and cannot be clearly resolved from the spectra after the outburst.
    
The $\gamma$-ray flux of \psr\ obviously decreased when the outburst occurred in 2016, but the spectral parameters are similar to those determined in the pre-outburst state.
The suppression of the radio emission was also detected at the outburst, and similar transitions of the radio flux for a pulsar were also in the mode changing RPPs.
Based on our studies, we find that after the outburst, the emission from the magnetosphere had a higher flux ratio ($\sim 30$\%) between the non-thermal and total flux of 3--65~keV if the contribution from the PWN did not have a significant change between epochs.
Such a high flux ratio contributed by the magnetospheric emission is also found to accompany with the active radio emission state. 
We also note that the pulsar had the largest spin-down rate between the end of 2016 Aug and early Sep when the radio emission of the pulsar was brightest \citep{Dai2018}.
An additional support of this picture in our analysis is that the magnitude of the frequency derivative reported by \citet{Archibald2016} is much smaller than the value we found before 2016 early Sep while the radio pulsar was still active.

\acknowledgments
%
This work made use of data supplied by the LAT data server of Fermi Science Support Center (FSSC) and by the archival data server of NASA's High Energy Astrophysics Science Archive Research Center (HEASARC).
This work is supported by the National Research Foundation of Korea (NRFK) through grant 2016R1A5A1013277.
H.-H.~W. and J.~T. are supported by National Science Foundation of China (NSFC) grants through grants 11573010, U1631103, and 11661161010.
C.-P.~H. and C.-Y.~N. are supported by a GRF grant of Hong Kong Government under HKU 17300215P.
C.~Y.~H. is supported by the NRFK through grant 2016R1A5A1013277.
A.~K.~H.~K. is supported by the Ministry of Science and Technology (MoST) of Taiwan through grants 105-2112-M-007-033-MY2, 105-2119-M-007-028-MY3, and 106-2918-I-007-005.
P.-H.~T.~T. is supported by NSFC through grants 11633007 and 11661161010.

{\it Facilities:} \facility{{\it Fermi}(LAT), {\it Swift}(XRT), {\it XMM}(EPIC), {\it NuSTAR}}. 



\end{document}